# APPLICATION OF GLOW DISCHARGE AES FOR INVESTIGATION OF METAL IONS AND WATER IN BIOLOGY AND MEDICINE.

Vasil G. Bregadze*, Eteri S. Gelagutashvili,  Ketevan J. Tsakadze.

Andronikashvili Institute of Physics, 6 Tamarashvili  st., Tbilisi, Georgia 0177;

∗Address for correspondence: e-mail :   **breg@iphac.ge**

**v_breg@yahoo.com**

Abstract

AES VHF inductively coupled plasmatron may be applied to wide range of studies. It enables rapid microanalysis of various solutions including biological objects and peripheral blood serum. In addition, it may be used for investigation of water desorption from solid bodies and for determination of energetic metal-macromolecule complexes.

Study of hydration energy and hydration number by kinetic curves of water glow discharge atomic spectral analysis of hydrogen (GD EAS analysis of hydrogen) desorption from Na-DNA humidified fibers allowed to reveal that structural and conformational changes in activation energy of hydrated water molecules increases by 0.65kcal/Mole of water.

The developed method of analysis of elements in solutions containing high concentrations of organic materials allows systematic study of practically healthy persons and reveals risk factors for several diseases. Microelemental content of blood serum fractions showed that amount of not bounded with ceruloplasmin copper was three times more at limphogranulomatose disease than that in healthy persons.

**Introduction**

It is difficult to overestimate the role of water and transition metal ions in the process of DNA functioning (super-spiral organization of nuclear DNA, replication, transcription, the translation of the genetic code, etc.) [1]. Especially interesting from this point of view is macro- and micro-elemental composition of real DNA molecules. There are many works [2-6] devoted to this problem.

The present paper represents the use of atomic-emission spectroscopy (AES) with very-high-frequency (VHF) Inductively Coupled Plasma(ICP) of reduced pressure for high sensitive and quick analysis of the trace quantities of elements in the samples of biological origin.

AES VHF inductively coupled plasmatron may be applied to wide range of studies. It enables rapid microanalysis of various solutions including biological objects and peripheral blood serum. In addition, it may be used for investigation of water desorption from solid bodies and evaluation of energetics of metal-macromolecule complexes.

The developed method of analysis of elements in solutions containing high concentrations of organic materials allows to provide systematic examination of practically healthy persons and to reveal risk factors of several diseases. Microelement content of blood serum fractions showed that amount of "free" – not bounded with ceruloplasmin Cu was three times more at limphogranulomatose disease than that in healthy persons.

## Materials.

Chemically pure chlorides of Mg, Mn, Co, Ni, Cu, Zn, Na and $Cr(NO_3)_3$, $AgNO_3$ and $NaNO_3$ were used. Bidistilled water served as solvent in all experiments. We studied calf thymus DNA (40% GC, "Serva") and DNA extracted from mice liver C3HA and ascetic hepatoma 22A in our Institute. The concentrations of nucleic acids were determined by UV absorption using molar extinction coefficient $\varepsilon$ =6600 $cm^{-1}M^{-1}$ at $\lambda$ =260nm. The double helix structure of the polymers was proved by their hyperchromic effect (>30%) and their typical thermal denaturation curve (in 0.01 M NaCl, pH 7.0); leucocytes were provided by the Institute of Hematology and Transfusiology of Georgian Ministry of Health; blood serum – from the polyclinic of Georgian Academy of Sciences; tumor tissues – from the Institute of Oncology of Georgian Ministry of Health;

## Methods

Method of multifunctional precise and consuming little energy atomic emission spectrometry (AES) is based on the plasmatron consisting of:-a disxharge tube piercing the inductive circuit of very high frequency (VHF) generator (110MHz) and connected with the gas-vacuum system maintaining low pressures (1-15Torr) of the flowing inert gas or ambient air; - a vaporizer for element analysis with electric system supporting the drying , ashing and transition of the sample placed on tantalum filament into aerosol state;-vaporizers used in isothermic and flash desorption regime; - a photoelectric spectrometer tuned to the spectral line of the analyzed element radiation in two regimes-in the pulse regime detection limits is 0.2-100 pg per absolute value;-in kinetics desorption regime the sensitivity is $10^{-6}$-$10^{-7}$ Torr for indication of partial pressure[11]. The main goal of the present work is to increase informational characteristics of glow discharge AES analytical method in combination with thermodynamic, hydrodynamic, molecular-spectroscopic and clinical methods of investigation (a) joint method of equilibrium dialysis and pulse AES allows to study stability of biopolymer-metal complexes. (b) the method combining water vapor desorption kinetics and AES analysis of hydrogen enables investigation  of hydration energy and hydration number of biopolymers. (c) the developed method allows routine microanalysis of copper in liquid mediums with high content of organic substances (including blood serum) that enables to reveal risk factors at prophylactic medical examinations.

**Element Analysis in High Concentrated solutions of Organic Matters.** The new method allowing to measure content of various microelements in the human peripheral blood serum is proposed. The basic idea of it is that after drying the process of ashing of the serum organic matter occurs in oxygen atmosphere in the quantities sparing the heated to 300-400°C tantalum filament of the vaporizer [12].

## Results and Discussions.

**Energetics of Metal-Macromolecule Interaction.** The energetics of complex formation can be investigated by equilibrium dialysis. From adsorption isotherms, we determine the binding (stability or association) constant, which is connected with the change of standard Gibbs free energy as follows:

$$\Delta G = -RT \cdot \ln K, \quad (1)$$

where K is the equilibrium binding constant, T − absolute temperature, R − gas constant equal to 1.9872 cal/deg.·mole.

Figure 1 shows the binding isotherm of Ni(II) ion with DNA from C3HA mice liver in the Scatchard coordinates. The points are experimental data and theoretical curve is received using Klothz method [13]. Use of different mathematical models reveals that two types of binding sites is the most accurate approximation. The results are presented in Table 2 (2mM NaCI, at room temperature).

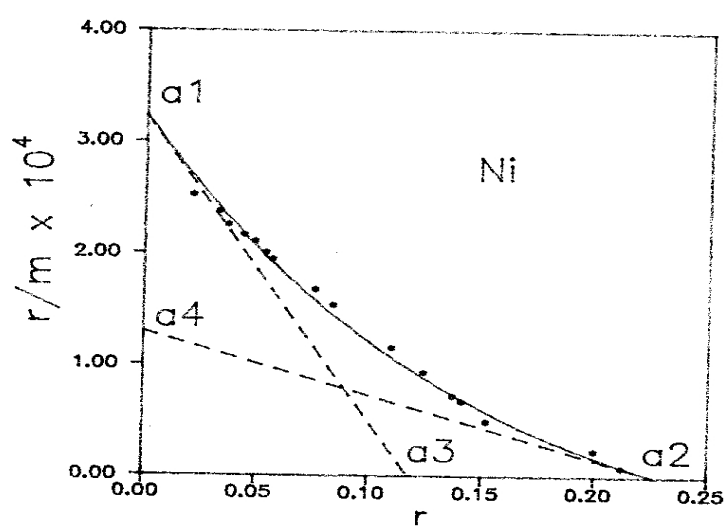

Fig.1. The binding isotherm of Ni(II) ion with DNA from C3HA mice liver
in the Scatchard coordinates.

The energetics of binding of Ni(II) and Zn(II) ions to DNA from C3HA mice liver and the ascetic hepatoma 22A cells were determined from their binding isotherms by equilibrium dialysis and pulse high frequency inductively coupled plasma atomic emission spectroscopy (Tab. 1).

Tab. 1. The energetics of binding of ions $Ni^{2+}$ and $Zn^{2+}$ to DNA from C3HA mice liver and the ascetic hepatoma 22A cells

| | | DNA, from C3Ha mice liver | | DNA, from ascetic hepatoma 22A | |
|---|---|---|---|---|---|
| Metal ions | | $Ni^{2+}$ | $Zn^{2+}$ | $Ni^{2+}$ | $Zn^{2+}$ |
| Microconstants | $k_1$x $10^4$ $M^{-1}$ | 30.4 | 20.7 | 36.9 | 20.6 |
| | $k_2$x104 $M^{-1}$ | 10.4 | 7.6 | 4.6 | 6.4 |
| Macroconstants | Kx $10^4$ $M^{-1}$ | 4.3 | 3.0 | 5.5 | 3.9 |
| Standard deviation | | 2.5 | 1.6 | 3.1 | 1.8 |
| | $\chi$ | 0.17 | 0.13 | 0.93 | 0.03 |

On average, every pair of phosphate groups is associated with a divalent metal ion in metal-DNA complexes at saturation in both cases. Investigation of interaction of the first row transition metals with normal and tumor DNA has revealed statistical discrepancy between $Ni^{2+}$ and $Zn^{2+}$ binding constants.

**The Investigation of Energetics of Water Molecule Binding with Biopolymers. Kinetics of Water Desorption from DNA.** In addition to elemental analysis, the low-pressure VHF plasma atomic-emission spectrometer enables watching of water desorption kinetics from surface of biopolymers [14]. All experiments were carried out at 20ºC in equilibrium conditions. DNA was humidified in advance at 25ºC using saturated solutions of $KNO_3$, $(NH_4)_2 \cdot SO_4$, NaCl, $NaNO_3$ and $Ca(NO_3)_2 \cdot H_2O$ which in hermetic volume provide 92.48%, 79,97%, 75,28%, 73.79% and 49.79% relative humidities respectively. The isothermal kinetic curve of water desorption from Na-DNA humidified at 92.5% (DNA mass is 2.74 mg) in relative units is given in Fig. 2.

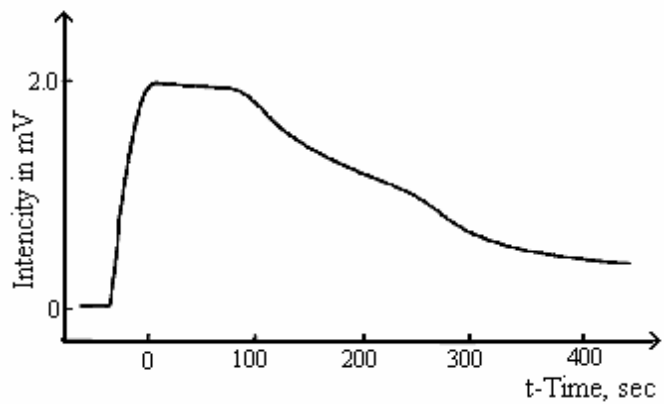

Fig.2. Kinetic curve of isothermal desorption of water from Na- DNA surface.
(on y-coordinate intensity of photosignal at $\lambda$=656.28nm (hydrogen line), which is proportional to the amount of water vapor passing through VHF plasmatron is plotted).

The desorption rate was determined from the curve plotted according to the Langmuir equation $\ln(M_e/(M_e - M_t))=kt$ in $\ln M_e/(M_e - M_t)$ and t coordinates, where t is desorption time, $M_e$ is amount of water, k is the desorption rate constant. The kinetic curve of water desorption from DNA in the solid state humidified at 92.5% is shown in Fig. 3.

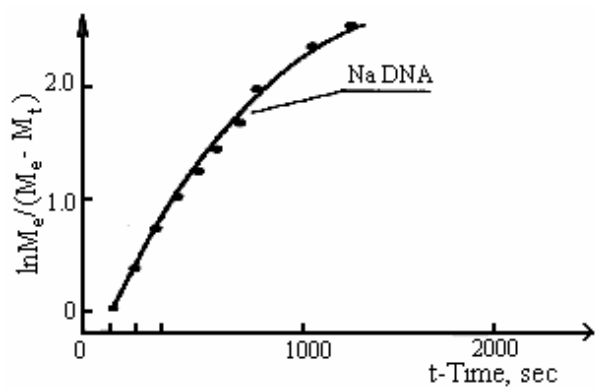

Fig.3. Kinetic curve of isothermal water desorption from
DNA surface    in Langmuir coordinates.

The study of kinetics of water desorption shows that at high humidity(60%) it follows the Langmuir law. The constant of desorption rate in the first order reactions is equal to

$$k_\alpha= \nu_o \exp(-E_d/RT), \qquad (3)$$

where $\nu_o$ is the exponential factor ($\nu_o= 10^{13}s^{-1}$), $E_d$ – energy of desorption activation. The desorption kinetics from humidified Na-DNA was studied at different relative humidities (Fig.4). The results show that structural and conformational changes in activation energy of hydrated water molecules increases by 0.65kcal/Mole of water.

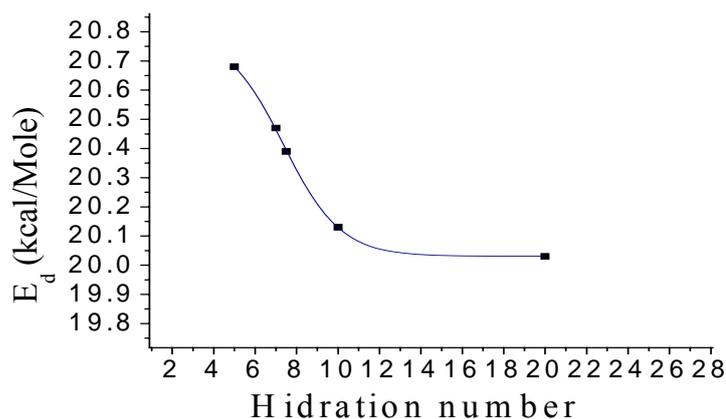

Fig.4. Influence of relative humidity on desorption energy of water from DNA surface.

This increase of energy equals to 0.1 kcal/Mole for B→C transition (from 20 to 10 water molecules per nucleotide) and 0.55 kcal/Mole for C→A →unordered state (from 10 to 5 molecules per nucleotide. This is one more indicative of importance of a hydration layer for DNA stability).

**Elemental Analysis in High Concentrated solutions of Organic Substances. Application in Medicine**. The developed method allows routine microanalysis of metals in liquid mediums with high content of organic substances (including blood serum) that enables to reveal risk factors at prophylactic medical examinations[12].

The histogram in Fig. 5 represents content of copper in blood serum of practically healthy persons (results of prophylactic check up). There were observed practically healthy 700 persons of the same social circle (research workers) from 25 to 50 years old.

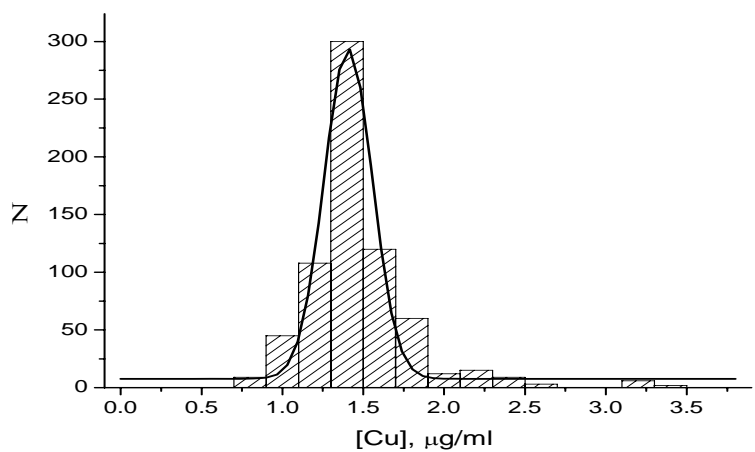

Fig.5 Distribution of Cu in the human peripheral blood serum.

The average value of Cu content was 1.34µg/ml, the standard deviation – 0,22µg/ml. As one can see, within the range $0.6 - 2$ µg/ml the distribution is well described by Gaussian law (the line in fig.5). However, there were observed certain number of cases with high content of copper beyond this distribution. Analysis of other blood characteristics of the same group revealed that those with high content of copper had increase hemoglobin and ESR (erythrocyte sedimentation rate). The erythrocyte and leukocyte amounts are not correlated with copper content (Tab.2).

Tab.2 Peripherial blood characteristcs  at different content of Cu in blood serum

| Cu (µg/ml) | | ESR | | | Hemoglobyn | | | Leucocytes | | | Erythrocytes | | |
|---|---|---|---|---|---|---|---|---|---|---|---|---|---|
| | | $\chi$ | $\Delta\chi$ | $\sigma$ | $\chi$ | $\Delta\chi$ | $\sigma$ | $\chi$ | $\Delta\chi$ | $\sigma$ | $\chi$ | $\Delta\chi$ | $\sigma$ |
| $0.6 - 2.0$ | | 8 | 0.67 n=25 | 3.35 | 76 | 0.88 n=67 | 7 | 5350 | 7,6 | 660 | 4,2 | 0.45 n=75 | 2,6 |
| $2.0 - 3.4$ | | 14 | 1.0 n=6 | 2.50 | 80 | 1.18 n=31 | 6.6 | 5300 | 95,4 | 530 | 4.8 | 0.10 n=32 | 2,5 |
| Statistical parameters | t | 5 | | | 2,74 | | | 0.006 | | | 1.13 | | |
| | p | 0.01 | | | 0.01 | | | - | | | - | | |

The further investigations showed that shift of the balance represented certain risk factor for different blood diseases. For instance, we found Cu content significantly increased in case of acute lymphatic leukemia (Table.3)

Tab. 3 Cu content in blood serum of patients with acute lymphatic leukemia

| N | 1 | 2 | 3 | 4 | 5 | 6 | 7 | 8 | 9 | 10 | 11 | 12 |
|---|---|---|---|---|---|---|---|---|---|---|---|---|
| Cu((µg/ml) | 2.7 | 2.3 | 2.4 | 1.6 | 1.8 | 1.9 | 1.8 | 2.0 | 1.7 | 1.6 | 1.8 | 1.7 |

In addition, redistribution of Cu in blood serum at lymphogranulomatosis was observed (fractionation was provided using Sephadex G-200 column) (Fig6).

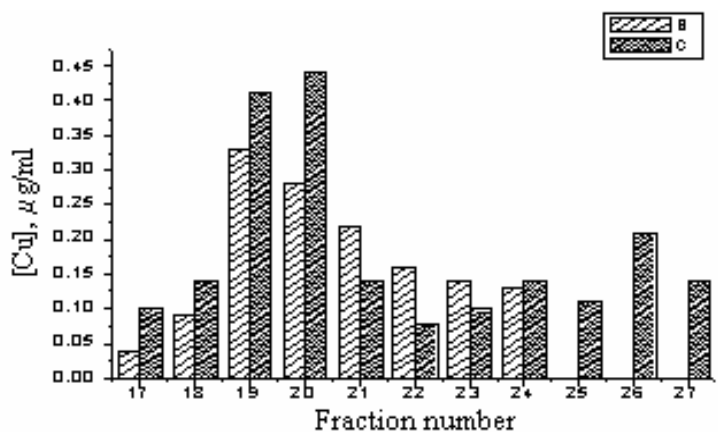

Fig. 6. Distribution of Cu in blood serum fractions of healthy persons ( series B) and at lymphogranulomatosis (series C)

Analysis of the data obtained in this case showed that although the total amount of copper in blood serum at this disease increases, their amount of bounded with ceruloplasmin is less then that in healthy persons. Amount of "free" – not bounded with ceruloplasmin Cu was three times more at limphogranulomatosis than that in healthy persons.

Copper was not the only metal showing different patterns in healthy persons and those with cancer disease. Table 4 compares endogenously bonded with DNA metals in non-malignant and cancerous cases of human breast tumor.

Tabl.4. Content of metals in human DNA (μg/g of dry mass) at non-cancerous and cancerous breast tumors. (The relative evaluating precision for Cd and Pb is 10%; in the rest of the cases – 5%).

| Metal | | Mg | Ca | Cr | Fe | Cu | Zn | Cd | Pb |
|---|---|---|---|---|---|---|---|---|---|
| DNA from Human breast tissue | Non-cancerous | 180 | 470 | 24 | 59 | 24 | 20 | 1.6 | 17 |
| | Cancerous | 200 | 580 | 30 | 76 | 50 | 24 | 2.5 | 23 |

## 8. Acknowledgments

The authors are very grateful to Prof, R. Abdushelishvili and Prof. N. Piradashvili, who have much contributed to this work with providing materials of clinical studies.

The authors thank Dr. A. Belokobilskii for extraction of DNA from different tissues.